%% file: main.tex
\documentclass[acmsmall]{acmart}

\AtBeginDocument{%
  }

\makeatletter
\def\@ACM@copyright@check@cc{}
\makeatother
\setcopyright{cc}
\setcctype{by-nc}
\acmJournal{PACMHCI}
\acmYear{2025} \acmVolume{9} \acmNumber{7} \acmArticle{CSCW446} \acmMonth{11} \acmPrice{}\acmDOI{10.1145/3757627}

\acmISBN{978-1-4503-XXXX-X/18/06}

\usepackage{multirow}
\newcommand{\revision}[1]{\textcolor{black}{#1}}

\begin{document}
\title[Mapping Community Appeals Systems: Lessons for Community-led Moderation \\ in Multi-Level Governance]{Mapping Community Appeals Systems: Lessons for Community-led Moderation in Multi-Level Governance}

\author{Juhoon Lee}
\email{juhoonlee@kaist.ac.kr}
\orcid{0000-0003-0796-4227}
\affiliation{%
  \institution{School of Computing, KAIST}
  \city{Daejeon}
  \country{Republic of Korea}
}

\author{Bich Ngoc Doan}
\email{ngocdb160900kaist.ac.kr}
\orcid{0009-0006-1767-2585}
\affiliation{%
  \institution{School of Computing, KAIST}
  \city{Daejeon}
  \country{Republic of Korea}
}

\author{Jonghyun Jee}
\email{jeejonghyun@kaist.ac.kr}
\orcid{0009-0009-7387-2508}
\affiliation{%
  \institution{Center for Digital Humanities and Computational Social Sciences, KAIST}
  \city{Daejeon}
  \country{Republic of Korea}
}

\author{Joseph Seering}
\email{seering@kaist.ac.kr}
\orcid{0000-0001-7606-4711}
\affiliation{%
  \institution{School of Computing, KAIST}
  \city{Daejeon}
  \country{Republic of Korea}
}

\renewcommand{\shortauthors}{Juhoon Lee, Bich Ngoc Doan, Jonghyun Jee, and Joseph Seering}

\begin{abstract}
  \input{sections/00_abstract}

\end{abstract}



\received{October 2024}
\received[revised]{April 2025}
\received[accepted]{August 2025}

\maketitle

\input{sections/01_intro}
\input{sections/02_rw}
\input{sections/03_methods}

\input{sections/04_results}

\input{sections/05_discussion}
\input{sections/06_conclusion}
\begin{acks}
We express sincere thanks to all community moderators who participated in our study for their valuable insights. This research was supported by the KAIST New Faculty Research Fund (Project No. G04230040: Building Integrated Models for Online Governance).
\end{acks}

\bibliographystyle{ACM-Reference-Format}
\bibliography{main}
\clearpage
\appendix
\input{sections/07_appendix}

\end{document}

%% file: sections/00_abstract.tex
Platforms are increasingly adopting industrial models of moderation that prioritize scalability and consistency, frequently at the expense of context-sensitive and user-centered values. Building on the multi-level governance framework that examines the interdependent relationship between platforms and middle-level communities, we investigate community appeals systems on Discord as a model for successful community-led governance. We investigate how Discord servers operationalize appeal systems through a qualitative interview study with focus groups and individual interviews with 17 community moderators. Our findings reveal a structured appeals process that balances scalability, fairness, and accountability while upholding community-centered values of growth and rehabilitation. Communities design these processes to empower users, ensuring their voices are heard in moderation decisions and fostering a sense of belonging. This research provides insights into the practical implementation of community-led governance in a multi-level governance framework, illustrating how communities can maintain their core principles while integrating procedural fairness and tool-based design. We discuss how platforms can gain insights from community-led moderation work to motivate governance structures that effectively balance and align the interests of multiple stakeholders.

%% file: sections/01_intro.tex
\section{Introduction}

Online governance has become a critical area of focus as digital platforms increasingly shape the dynamics of community interaction and engagement. Moderation has often been described as a balancing act, requiring a choice between scalable, rule-based decisions and local, context-sensitive approaches~\cite{ehrett2016ejudiciaries, fan2020juries, lampe2004slashdot, kayhan2013content, seering2020reconsidering, helberger2018cooperative}. While community-driven moderation emphasizes norm-setting and accountability, enabling users to shape the standards of their interactions, the expansion of safety-focused regulation has pushed platforms to emphasize compliant, industrialized governance to meet commercial and legal obligations, which risks creating a rigid system that forgoes contextual nuance~\cite{keller2024compliant}. This raises the question: Is online governance doomed to be a regulation-driven battle for control between platforms and communities?

The multi-level governance framework proposed by Jhaver et al. provides a promising alternative to this dichotomy~---~which positions platforms and users against each other~---~by emphasizing complementary interactions among the units of governance at different levels\revision{~\cite{jhaver2023multilevel, forte2009decentralization, hasinoff2022subsidiarity}}. Jhaver et al. specifically highlight the value of middle-level management practices that cater to the unique needs and norms of individual communities~\cite{jhaver2023multilevel}. However, creating a governance structure that integrates regular users, platform policymakers, community moderators, and their commercial counterparts in a coherent architecture is no simple task.

This study examines community appeals systems in Discord as a compelling example of how the type of fluid, context-specific judgment that is increasingly absent from compliance-focused platform processes can be reintroduced at the community level when users are given sufficient means to do so. \revision{A \textit{community appeals system} refers to a systemized process through which a community member requests community moderators to review and potentially overturn moderation decisions. It allows users to challenge enforcement measures such as bans and message deletions in a structured manner governed by the community’s own rules and standards.} An appeals system enforces accountability on authorities by empowering users to contest moderation decisions~\cite{vaccaro2021contestability, vaccaro2020contesting}. While platforms typically use appeals systems to correct erroneous moderation actions, communities adopt a rehabilitative approach, using appeals to reintegrate users and uphold social norms. However, this comes at the cost of increased strain on moderators. In turn, prior research by Atreja et al. designed a system to mitigate the emotional toll and workload faced by moderators in appeals systems on Reddit~\cite{atreja2024appealmod}. Similarly, Discord communities have organically leveraged the technical affordances provided by the platform to achieve supportive, context-sensitive moderation across millions of communities.\footnote{\url{https://discord.com/company}} \revision{We extend upon a series of works by Vaccaro et al. on the challenges and desires of users using existing appeals paradigms by exploring how Discord community appeals systems manage such tensions.}

By examining the implementation of Discord community appeals processes in detail, we aim to contribute insights into the effective design of community-led governance structures as well as the value these processes bring to communities. We conducted an interview study with 17 Discord moderators, mapping their servers' appeals processes from initial review to reintegration. We first conducted focus group interviews with three groups of three moderators operating appeals systems to understand the general structure of Discord appeals systems and their benefits, then narrowed down the probe on each step, detailing the evaluation and collaboration processes through individual interviews (N=8). We then performed an inductive thematic analysis to identify the core steps. We contribute a structured process map of the appeals workflow to visualize these findings~\cite{nelson2011quality}. 

We answer the following research questions:
\begin{itemize}
  \item RQ1. How do community moderators design appeals systems to fit the needs of their communities?
  \item RQ2. How do community moderators ensure procedural fairness while handling appeals at scale?
  \item RQ3. What values do appeals systems bring to communities and platforms?
\end{itemize}

Our findings showcase how a community can design and implement streamlined moderation decision-making systems that cater to the resources, characteristics, and norms of the community. We observe that moderators define judicial and technical processes to maintain fairness and accountability throughout the appeals process. Most importantly, Discord appeals systems illustrate how community managers can systematically incorporate rehabilitative principles while maintaining scalability. This study contributes to the growing body of work demonstrating the importance of decentralized governance helmed by middle layers of governance. We discuss how mapping and analyzing local, community-driven processes can support platform governance in adopting a more flexible and context-aware approach to moderation.

%% file: sections/02_rw.tex
\section{Related Work}
Community appeals systems refer to the specific system in which a community allows its users to contest moderation decisions. While we probe and map this process thoroughly in this work, we place community appeals systems in a larger framework of online governance -- specifically, their values and implementation from a multi-level governance perspective. We first describe the limitations of existing online governance frameworks and the contributions of community-led governance in a decentralized, multi-level governance framework. Then, we explore how community moderators adopt tools to support community moderation tasks. Finally, we show how community appeal systems contribute to the growing literature on realizing rehabilitative justice in moderation. 

\subsection{Community Moderation in Multi-Level Governance}
Historically, perspectives on online governance have often been presented from one of two main perspectives: platform-driven governance, where platform authorities impose top-down rules~\cite{gillespie2018custodians, caplan2018content, chen2021decentralized}, and community-driven governance, where local users or smaller groups within the platform establish and enforce their own norms~\cite{seering2020reconsidering, grimmelmann2015virtues, matias2019civic, seering2019moderator}. Platform-driven governance involves centralized decision-making guided by a company's terms of service, content policies, and the algorithms underlying automated moderation tools~\cite{caplan2020demonetization, roberts2016commercial}. On the other hand, community-driven governance tends to be more decentralized and participatory, allowing users to self-organize, often in a form where communities are led by volunteer community moderators who enforce their own standards of conduct~\cite{seering2019moderator, fiesler2018reddit}. The two structures may see overlap in the content they moderate but have generally been treated as if they are operating at fundamentally different scales and purposes. 

This separated, binary view of governance limits the potential for complementary or even cooperative interactions between the platform and its communities in an increasingly rigid, regulation-driven content moderation environment. Works on moderation have often identified tensions and trade-offs between moderation styles that place platform and community moderation at opposite ends\revision{~\cite{ehrett2016ejudiciaries, kayhan2013content, seering2020reconsidering} across different models~\cite{fan2020juries, lampe2004slashdot}, platforms~\cite{gilbert2020cesspool}, and mechanisms~\cite{helberger2018cooperative} of governance}; for example, prior research has argued that platform-driven moderation pursues efficiency through automation and increased engagement, while distributed or community-driven governance promotes norm-setting and accountability~\cite{jiang2023tradeoff}. Similarly, in Caplan's taxonomy of platforms' approaches to content moderation, ``community-reliant'' approaches are seen as fundamentally separate from ``industrial'' approaches, with the former more sensitive to local context and the latter more consistent and efficient at scale~\cite{caplan2018content}. The existence of such a contrast implies that platforms operating more small-scale, in-house moderation can evolve into more sophisticated, industrial models as their business interests change. Increasing pressure from regulatory compliance may be forcing this transition; per Keller, the need to present an auditable content moderation process in the face of regulations such as the EU's Digital Services Act (DSA) is pressuring platforms to adopt methods that are inflexible and less sensitive to contextual nuance~\cite{keller2024compliant}. Framing platform- and community-driven approaches to governance as fundamentally separate models pits them against each other, rather than allowing them to integrate where, for example, the attention to nuance in one approach could complement the efficiency and scalability of the other. 

In response to these limitations, scholars have proposed the adoption of a multi-level governance framework that enables a more holistic understanding of online governance~\revision{~\cite{jhaver2023multilevel, wu2023colorblind, forte2009decentralization, hasinoff2022subsidiarity}}. Jhaver et al. defined online multi-level governance as decentralization through the vertical and horizontal interplay between the centers of power in an online platform, including companies, community leaders, and users~\cite{jhaver2023multilevel}. Under this framework, governance is less the outcome of either platform or community initiatives but more the result of interactions in authority and autonomy between these layers. Their work particularly emphasizes the contributions of ubiquitous middle levels of governance (such as community managers, third-party tool developers, and server admins) as key mediators between top-down policies and grassroots community norms. However, though Jhaver et al. showed that a multi-level platform structure has advantages in governing in a way that is sensitive to the interests of a broader variety of users, designing a coherent, planned, integrated multi-level governance architecture is no easy task.

\revision{The task is especially difficult when one must consider the contextuality of the culture, norms, and identities unique to the community. The role of cultural context in content moderation and platform governance has become increasingly important as digital platforms grow globally. Western-centric models of content moderation are often misaligned with norms and accepted behaviors of diverse, international user bases~\cite{shahid2023decolonizing}. These models tend to overlook the complexities of cultural differences, leading to a misalignment between platform policies and the cultural values of users from non-Western contexts, particularly in the Global South where ``sociotechnical systems reinstate colonial structures and values''~\cite{das2023marginalization}. The prevailing emphasis on universal governance structures often disregards the diverse and sometimes conflicting local norms of its users, which can significantly influence how users understand fairness, justice, and responsibility within online spaces.}

\revision{Incorporating cross-cultural perspectives is essential for creating more effective and equitable content moderation systems. Many digital governance models fail to account for the varied cultural practices and expectations that shape how communities interact with platforms and one another. For instance, Garimella demonstrated that group norms and the degree of familiarity among members in WhatsApp groups influenced the patterns of community-driven fact-checking~\cite{garimella2022}. Different cultures may also evaluate governance structures differently; Graeber and Wengrow emphasize that the idea of democracy being inherently superior as a form of governance is rooted in ``Western'' cultural influence and should not be taken as the universal ideal~\cite{graeber2021dawn}. In this way, the structure of online governance must be able to reflect the unique identities and needs of its members.} 

Among the many processes identified within community-driven governance, community appeals systems offer a unique window into how localized decision-making can complement overarching governance frameworks. Unlike platform-driven appeals processes -- such as the TikTok content violation appeals~\cite{tiktok2024content} or YouTube’s tiered appeals process~\cite{caplan2020demonetization} -- which aim primarily to overturn errors, community appeals processes delve deeper into social dynamics~\cite{vaccaro2020contesting}, allowing for more personalized and context-sensitive judgments. Community moderators use these appeals not simply to correct moderation ``mistakes'' but to assess a user's readiness to reintegrate and contribute positively to the community~\cite{atreja2024appealmod}. This distinct function highlights the relational, rather than purely procedural, purpose of community-led appeals. In this work, we specifically investigate the appeals process within Discord servers, which serve as successful examples of systematic community-driven governance. Discord community appeals systems exemplify the possibility of a balanced approach -- one that merges the structure and scale of platform policies with the empathy and local awareness inherent to community-driven moderation. Discord server moderators have created a structured process that encourages collaboration and implementation of technical tools to streamline moderation and enhance fairness in decision-making.

\subsection{Community Governance Tools for Moderators}
Though community moderation remains a fundamentally human-centered process, moderators have benefited from the development of a range of innovative tools that facilitate the management and enforcement of community standards. These tools include \revision{automated moderation systems in the form of filters~\cite{jhaver2022designing}, detection systems~\cite{haque2022citadel, yin2009detection}, automoderator bots~\cite{kiene2019frames, jhaver2019automoderator, chandrasekharan2019crossmod}}, \revision{community-driven~\cite{wang2024efficiency, jhaver2018online, rifat2024islamophobia}} or \revision{``friendsourced'' moderation tools~\cite{mahar2018squadbox}}, and  
even proactive support mechanisms~\cite{schluger2022proactive, kriplean2012considerit, kriplean2012reflect, kim2024respect}.
Automated moderation systems that detect and remove harmful content are often the first line of defense against harmful content, designed to scale quickly and efficiently. They perform tasks such as filtering harmful content~\cite{jhaver2022designing, chandrasekharan2019crossmod} and detecting hate speech~\cite{gunturi2023toxvis}. Examples include tools like CrossMod, a Reddit bot that predicts which comments moderators are likely to remove through cross-community learning~\cite{chandrasekharan2019crossmod}.
However, due to their automated nature, these systems sometimes misclassify content or overreach, leading to erroneous and biased penalties or the removal of legitimate content~\cite{gomez2024arbitrariness, song2023modsandbox, thiago2021fighting, binns2017bias}. Moreover, over-delegation of decisions to bots may lead \revision{community members to experience a sense of reduced agency over the governance of their communities~\cite{heer2019agency, cai2019imperfect, vaccaro2021contestability}, especially when transparency is lacking behind the decision-making~\cite{suzor2019transparency, ma2023experience, jhaver2019did}}. 

Communities address these limitations by often supplementing automated moderation with community-driven tools, which empower users to participate actively in governance by flagging or reporting problematic content~\cite{hee2024brinjal, jia2022misinformation}. Other works integrate social support into moderation systems~\cite{blackwell2017heartmob, yu2020care, mahar2018squadbox}. For example, Squadbox enables users to manage harassment by requesting moderation help from friends~\cite{mahar2018squadbox}. These decentralized, user-centered tools foster a sense of ownership and accountability within communities, reinforcing their capacity for self-regulation. Other tools provide more administrative or logistical support for moderators in human-driven processes through enhanced analytics and decision-support systems, such as visualization of content trends and potential rule-breaking behaviors~\cite{choi2023convex, park2016commentiq, song2023modsandbox}. Recent works have also emphasized proactive measures: Chillbot allows moderators to safely intervene before a situation escalates into harm by sending feedback responses~\cite{seering2024chillbot}. With the increasing availability of more and more sophisticated moderation tools, moderators are tasked with selecting the most effective approaches to support their communities.

Despite the challenge of selecting appropriate tools from a vast ecosystem~\cite{geiger2010work, jhaver2019automoderator}, many moderators have successfully integrated tools to aid and shape social processes within their community~\cite{cai2024thirdparty, hsieh2023bots, hwang2024bots}. Using third-party bots or building custom tools demands significant technical skills and effort, but many communities on platforms such as Discord value the development of custom tools highly enough to dedicate significant resources toward building customized, user-centered tools through collaborative and communal effort~\cite{hwang2024bots}. Discord users have cultivated a thriving marketplace for exchanging skills and knowledge related to both commercial and open-source moderation bots, such as YAGPDB,\footnote{https://yagpdb.xyz/} MEE6,\footnote{https://mee6.xyz/} Carlbot,\footnote{https://carl.gg/} and Dyno.\footnote{https://dyno.gg/}

In this work, we explore Discord community (i.e., server) appeals systems to understand how moderators effectively apply tools to support local, context-aware governance. As we discuss later in this paper, the community appeals systems in Discord make use of both automated and participatory moderation tools to handle a high number of cases with attention to context and interpersonal dynamics. Compared to other moderation tasks, handling appeals requires a deep understanding of the context behind often sensitive situations. When a moderator reviews possibly harmful and toxic content, they must perform intensive cognitive and emotional labor~\cite{dosono2019emotional, gilbert2020cesspool, schöpkegonzalez2024quit, steiger2021psychological}. To aid moderators in this work, Atreja et al. created AppealMod, an appeals process support system for Reddit designed to increase efficiency for moderators and reduce their exposure to harmful content~\cite{atreja2024appealmod}. AppealMod mitigates the imbalanced workload between a Reddit moderator and the user by requiring the user to fill out detailed information. We build on this work to showcase how Discord moderators use tools to develop a systematic process that not only improves efficiency but also upholds values of due process, fairness, and rehabilitation. 

\subsection{Rehabilitation through Contestability}
Community appeals systems show promise for fostering healthier communities in alignment with principles of rehabilitation in restorative justice models. Traditionally, online moderation has often been punitive, focusing on sanctions such as content removal, suspensions, or bans to maintain order~\cite{jhaver2019did, grimmelmann2015virtues}. Even when moderation is not explicitly designed to be punitive, platforms' prioritization of standardized and scalable content-based moderation practices~\cite{caplan2018content} leads to a result that is far from restorative. These industrial approaches may be effective in removing immediate harm, but they often fail to address underlying issues, potentially leading to recurring problems~\cite{hasinoff2020restorative}.

In response, recent works have applied the framework of restorative justice to moderation practices~\cite{petersonsalahuddin2024repairing, xiao2023restorative, hasinoff2022subsidiarity, kou2021punishment, schoenebeck2021justice}. In contrast to punitive or retributive justice models, restorative justice emphasizes healing and reintegration, aiming to repair the harm by putting the focus on the individuals rather than the content of the harm~\cite{wenzel2008retributive, ma2023transparency}. Restorative justice systems embrace three key principles: (1) recognizing and responding to the victim's needs arising from the harm, (2) encouraging rehabilitation through accountability and restoration of harm by the perpetrator, and (3) involving the community to foster collective support and reconciliation~\cite{mcccold2000holistic, zehr2015little}. Restorative justice practices have shown effectiveness in reducing repeat offenses and repairing harm caused~\cite{latimer2005effectiveness}. But despite the benefits, implementing restorative strategies in real-life online contexts has proven to be difficult when the positive impact of the strategies was not immediately and easily recognized by the stakeholders~\cite{xiao2023restorative}. 

The real-life examples of community appeals processes embody the principle of rehabilitation in restorative justice by introducing contestability. Contestability is recognized as a key principle of ethical conduct in online moderation~\cite{santaclara}. This interaction can serve as a learning experience, where users gain a clearer understanding of the rules they violated and the impact of their actions on the community. For example, platforms may require users to explain their reasoning or acknowledge the harm caused by their actions as part of the appeals process~\cite{atreja2024appealmod}. In doing so, users can reflect on how they can modify their behavior moving forward. This approach in community appeals systems is closely tied to the concept of reintegration of the perpetrator, a core tenet of rehabilitative and restorative justice~\cite{cullen2013reaffirming, seigert2007rehabilitation, kou2024peer}. In this work, we examine how communities offer a more compassionate and constructive form of governance through appeals.

%% file: sections/03_methods.tex
\section{Methodology}
To understand how moderators handle appeals and to produce a generalized process map for community-driven appeals, we conducted semi-structured \revision{interviews} with Discord moderators in servers with existing appeals systems. The interviews were conducted in two phases: We first interviewed focus groups to compare the general experiences and perceived values of handling the appeals process. This methodology follows the procedures of Singh et al., who leveraged focus group interview results to identify thematic differences in cyberbullying experiences on social platforms, subsequently guiding individual interviews on more personal encounters with specific app features~\cite{singh2017cyberbullying}. Afterward, we interviewed individual moderators to analyze in-depth how each moderator evaluates an appeal, how tools support the appeals process, and the mechanics of the collaborative decision process. This study was reviewed and approved by the Institutional Review Board (IRB) at KAIST. 


\subsection{Participants}
We recruited the participants through moderator community servers in Discord and via \revision{snowball sampling~\cite{atkinson2001accessing}}. We first posted the focus group interview recruitment post on a community server where Discord moderators gather to discuss issues related to moderation, which outlined the goal of the research and requested moderators who had experience in running an appeals system for their servers. For individual interviews, we also reached out to moderators who had participated in a previous study related to community moderation and who had mentioned having appeals systems in the servers they moderated. We increased diversity in the interview pool by selecting moderators from servers with varying community characteristics, such as server topics. We also verified that all moderators were moderating one or more servers with more than 1,000 members and were referring to different servers when describing their experiences with appeals systems. 

We conducted semi-structured interviews with three randomly assigned focus groups of three moderators each (N=9 total interviewees). For individual interviews, we recruited a total of eight additional moderators. Detailed information on the servers, including the type of appeals systems and tools used, is outlined in Table \ref{table:participant}. We did not collect any additional demographic information on the moderators as we aimed to minimize personal information shared to respect the moderators' privacy.

Two of the authors conducted the focus group interviews, while the individual interviews were conducted by the first author. The interviews were mostly conducted via Discord voice calls. We used text-based channels for one focus group interview (FG1) and one individual interview (P2) due to the moderators' accessibility-related requests. \revision{Though online text-based interviews often result in a smaller volume of data, previous research has shown that both synchronous and asynchronous text-based groups can offer interaction levels that are comparable to in-person groups~\cite{jones2022focus}. Text-based group and individual interviews show no qualitative differences in thematic content when compared to other in-person and online modalities~\cite{guest2020focus}. We found that while the participants of text-based focus groups tended to give longer answers, the participants engaged with other participants' answers in their responses. A challenge of text-based interviews was maintaining the context and momentum of the discussion, as the interview spanned multiple days. We mitigated this by asking detailed questions, directly referencing answers given by participants, and pinging participants if the conversation paused for too long.} All voice-based interviews lasted approximately an hour, while the two text-based interviews concluded within a week of the first question being asked. Participants were compensated with a gift card worth 25 USD converted to the type and currency of their choice for completing the interview.

\begin{table}[]
\resizebox{\textwidth}{!}{%
\begin{tabular}{@{}cclll@{}}
\toprule
\textbf{Session}            & \textbf{Participant ID} & \multicolumn{1}{c}{\textbf{Server Topic}} & \multicolumn{1}{c}{\textbf{Appeals System Type}} & \multicolumn{1}{c}{\textbf{Tool(s) used}} \\ \midrule
\multirow{3}{*}{FG1}        & F1                      & Sports                                   & Appeals server                                   & General-use moderation bot                \\ \cmidrule(l){2-5} 
                            & F2                      & Gaming                                   & Appeals server                                   & ModMail                                   \\ \cmidrule(l){2-5} 
                            & F3                      & Culture and language                     & Appeals server                                   & Ticket Tool, General-use moderation bot   \\ \midrule
\multirow{3}{*}{FG2}        & F4                      & Gaming, Content creator                  & Appeals server                                   & ModMail, Zeppelin                         \\ \cmidrule(l){2-5} 
                            & F5                      & Gaming                                   & Appeals server                                   & ModMail                                   \\ \cmidrule(l){2-5} 
                            & F6                      & Software development                     & Email                                            & General-use moderation bot                \\ \midrule
\multirow{3}{*}{FG3}        & F7                      & Fandom                                   & Google Form                                      & General-use moderation bot                \\ \cmidrule(l){2-5} 
                            & F8                      & Gaming                                   & Appeals server                                   & General-use moderation bot                \\ \cmidrule(l){2-5} 
                            & F9                      & Gaming, Fandom                           & Appeals server                                   & General-use moderation bot                \\ \midrule
\multirow{8}{*}{Individual} & P1                      & Content creator                          & Bot-based form                                   & ModMail, Carlbot, MEE6                    \\ \cmidrule(l){2-5} 
                            & P2                      & Content creator                          & Bot-based form                                   & Dyno                                      \\ \cmidrule(l){2-5} 
                            & P3                      & Gaming, Content creator, Hobby           & Direct Messaging                                 & General-use moderation bot                \\ \cmidrule(l){2-5} 
                            & P4                      & Gaming, Content creator                  & Bot-based form                                   & Dyno                                      \\ \cmidrule(l){2-5} 
                            & P5                      & Content creator                          & Appeals server                                   & Sapphire                                  \\ \cmidrule(l){2-5} 
                            & P6                      & Gaming                                   & Appeals server                                   & ModMail, Dyno, Zeppelin                   \\ \cmidrule(l){2-5} 
                            & P7                      & Content creator                          & Appeals server                                      & BlueBot, Zeppelin                         \\ \cmidrule(l){2-5} 
                            & P8                      & Gaming                                   & Appeals server                                   & Dyno                                      \\ \bottomrule
\end{tabular}
}
\caption{Demographics of moderator participants and their servers. \textit{Server Topic} refers to the broad topic of the communities, such as ``Fandom'' for a community based around TV shows. Some participants mentioned multiple appeals systems across several servers. \textit{Appeals System Type} refers to the primary method in which appeals are received and handled. \textit{Tool(s) used} refer to specific third-party tools used by moderators in the appeals process. If the participant mentioned a moderation bot but did not name a specific bot framework, we denote it as a ``general-use moderation bot''.}
\label{table:participant}
\end{table}

\subsection{Focus Group Interviews}
The focus group interviews gathered three groups of three moderators to share and discuss their experiences with handling appeals and the value they saw in having an appeals system for their communities. We chose to begin the probe with focus group interviews as it is important to not only recognize the commonalities in the appeals experiences but also to compare why and how diverse appeals systems take form according to server needs. Additionally, conducting focus group interviews helped interviewees encourage each other to provide opinions on novel points mentioned by another moderator or jog their memory on similar aspects of their servers. We note that focus group interviews have been shown to stimulate deeper insights by having participants build upon or challenge others' ideas~\cite{lampinen2011interpersonal,
wisniewski2012boundary}, serving as a useful method for expanding the initial understanding of a specific process.

The focus group interviews began by first introducing the goal of the study and requesting participants to be respectful towards other participants, with a focus on maintaining anonymity and withholding personal information when sharing anecdotes. The participants were encouraged to jump in and engage with other participants, responding to or adding on to their ideas, asking questions, or pointing out differences. The participants took turns introducing themselves to others and summarizing their general moderator experience, noting the types of servers they moderate. Afterward, one of the authors began the discussion by first asking about the appeals cases the moderators face in their servers (e.g., ``What types of appeals cases have you seen?'', ``Can you recount a typical appeals case?''). The moderators were then asked to describe the process of how an appeal is typically handled in their server(s) (e.g., ``What does the appeals process look like?'', ``What are the main characteristics of a server that affect what kind of appeal systems you use?''). Finally, the participants were asked to discuss among themselves the value of maintaining an appeals system.

\subsection{Individual Interviews}
The research team thematically analyzed the results to identify gaps in our understanding of critical steps and components of the appeals process. The analysis of the focus group interviews is explained in detail in the next subsection. In doing so, we recognized that while the focus group interviews touched on a number of core themes, there was more to be learned specifically about the process for evaluating appeals, the collaboration dynamics between multiple moderators, and the implementation details, such as how tools are incorporated into the design of processes. Thus, we conducted semi-structured individual interviews with eight additional moderators to further probe each individual moderator's judgment protocols and to dive into the design details of individual servers. In these interviews, some participants voluntarily chose to share their screens to demonstrate in real-time the inner workings of the appeals system. We provide the structured interview questions for both focus group and individual interviews in Appendix \ref{A}.

\subsection{Thematic Analysis}
After the focus group interviews, we conducted an inductive thematic analysis using the method outlined by Braun and Clarke~\cite{braun2006thematic} to identify the key components of the appeals process and the value it brings. The authors transcribed the interviews for analysis. Three of the authors generated the initial codes by individually performing line-by-line open coding on all three focus group interviews. Next, the authors discussed and combined the preliminary codes into an initial codebook. We identified 36 initial codes through this process. The first author re-coded the interviews with the codebook, checking its completeness and validity. Afterward, we categorized similar codes into themes, organizing based on the characteristics of the appeals system design, the steps in the appeals decision-making process, collaborative dynamics, and the value incurred by having an appeals system.

After the completion of the subsequent individual interviews, the text was transcribed by the authors. If the participant had shared their screen, the researchers tagged the screenshots to the places in the transcript where they were referenced. Three of the authors performed the same analysis process as described above for the focus group interviews on two of the individual interviews. Through discussion, the authors added, edited, and merged codes into the intermediate codebook, resulting in a total number of 43 codes that fortified the themes of evaluation and collaboration processes. The first author coded the remaining individual interviews, developing 45 codes in total. The final codebook was then validated by the second and third authors.

\subsubsection{Process Mapping}
We constructed a \textit{sequential process map} to visualize how appeal cases flow through the system. Sequential process maps have been prevalent representations of processes in fields such as management and healthcare~\cite{antonacci2021process, mandelburger2013management, colligan2010process}. We used the interview results to construct a sequential flow diagram of appeal processes based on the methods described by Nelson et al~\cite{nelson2011quality}. We first determined the boundaries of the process based on our aim statement of understanding how an appeal will be processed by community moderators. We then defined high-level process steps of the appeals process (e.g., initiate an appeal, gather information, collaborate on the appeal), then broke down the large steps into smaller activity steps (e.g., give out initial enforcement, check time since the appeal, discuss points of contention). The first author created the initial draft of the diagram. Then, the second and third authors checked the diagram and gave feedback on action granularity and importance, connections and relationships, and readability. The authors iterated through the process until no further suggestions for modifications were given. The final representation of the process map is seen in Figure ~\ref{fig:map}.

\revision{\subsection{Positionality Statement}}

\revision{As researchers studying online moderation processes, we reflect on our position in this work. The authors of this paper seek to contribute to the growing body of work on achieving effective models of online moderation through collaborations between diverse stakeholders across all levels of governance. We leverage our personal experience as Discord users, with all authors being members of different servers spanning several years, to examine the implications of polycentric governance within the familiar infrastructure of Discord servers. Some of the authors have actively explored and engaged with Discord’s moderation ecosystem through research in the past, working directly with platform employees to understand community moderation practices and building tools supporting Discord moderators. We also recognize that our identities --- including but not limited to demographic identities, cultural backgrounds, and institutional affiliations --- affect our views toward how online platforms should be moderated and governed. Our perspectives may differ from those of Discord users who experience moderation decisions firsthand, particularly those from marginalized or underrepresented communities. We recognize that moderation is deeply influenced by context and the evolving needs of diverse communities and the process of evaluating these systems must be continually assessed and redefined.}

%% file: sections/04_results.tex
\section{Results}
\begin{figure}
    \centering
    \includegraphics[width=1.0\textwidth]{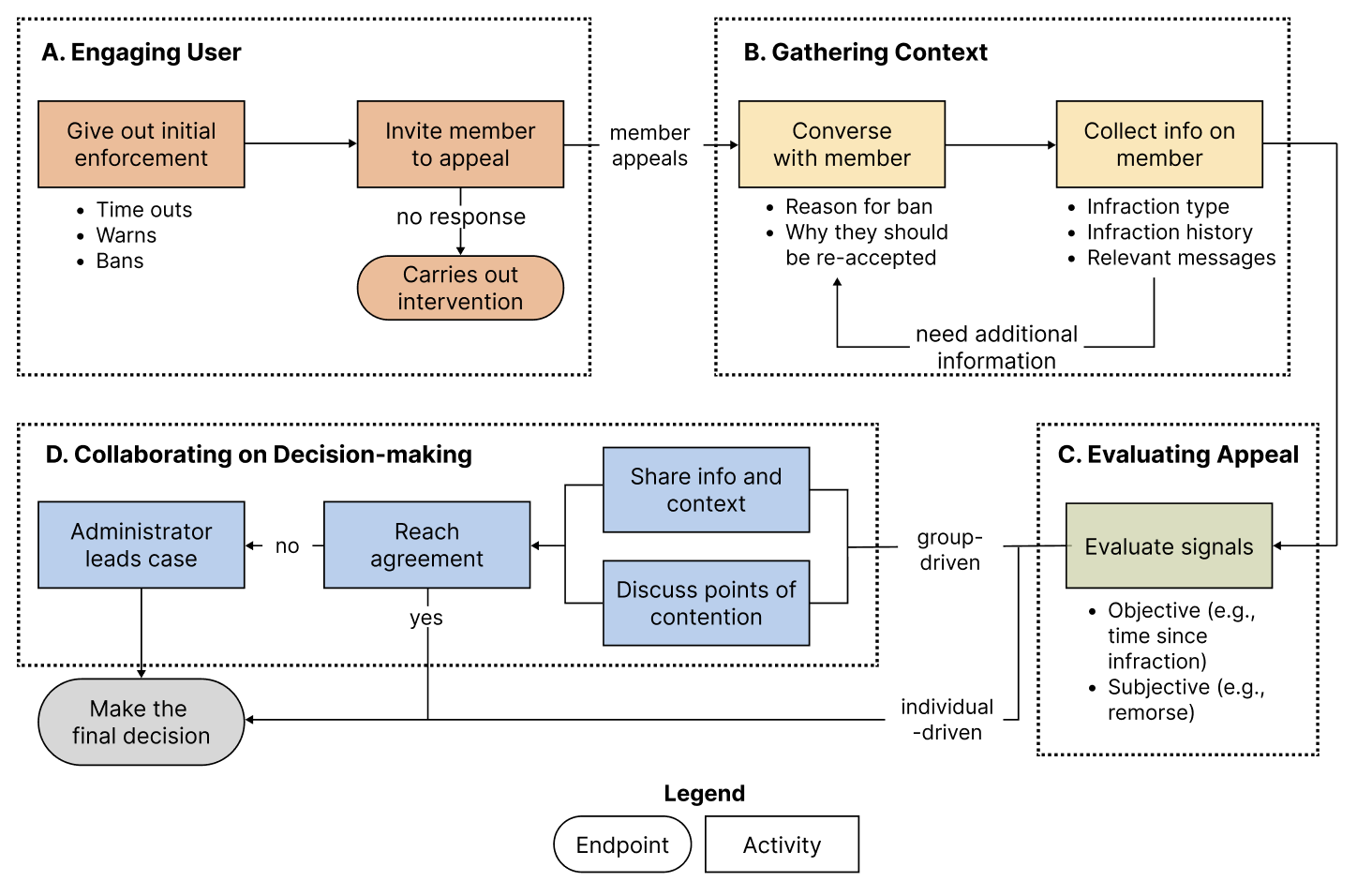}
    \caption{The sequential process map of the community appeals process in Discord servers by community moderators. The dotted boxes outline high-level processes, with the detailed activity represented in a rectangular box.}
    \label{fig:map}
\end{figure}

We describe how Discord communities carry out appeals processes in detail based on the study results, showcasing the actions taken to ensure a timely, thorough, and fair evaluation process. We structure our results in line with the process map of appeals systems illustrated in Figure ~\ref{fig:map}, highlighting the process design decisions taken by community moderators at different stages of the appeals process. Finally, we address the challenges and value of operating an appeals process at the community level.

\subsection{Initiating an Appeal}

\subsubsection{Engaging the user}
When a community member violates the rules in a Discord server, moderators often enforce interventions in the form of warnings, timeouts, or bans, with a determination made based on the severity of the infraction (Figure ~\ref{fig:map}A, left). When asked about the common types of violations that they face when handling appeals, moderators mentioned interpersonal conflicts, slurs and hate speech, hacked accounts, offensive content, and content spamming as some of the most common cases. The distribution of appealed infraction types differed based on the topic and demographic of the server -- for example, in P7's server for a YouTube content creator whose content was sometimes considered controversial and whose fanbase was mostly adolescents, slurs contributed to a large number of infractions that would be appealed. The young members would appeal their timeouts or bans as they often lacked understanding of why it would be considered problematic.

The differences in server characteristics also shape how the appeals are instigated. When a member is notified that they have been punished for a rule infraction, they are also sent instructions on how they can appeal the moderation decision, usually through Direct Messaging (DM) via a moderation bot (most typically ModMail (Figure ~\ref{fig:map}A, right).\footnote{https://modmail.xyz/}) Many moderators mentioned difficulties in reaching the member at the initial stage. Users may not always check their messages or may simply give up rejoining a server after being banned. Typically, smaller servers with a more tightly-knit community and fewer moderators used DMs, emails, or Google Forms to receive appeals. Larger servers with hundreds of thousands of members required a much more streamlined and systematic appeals system to handle the large volume of appeals and to acquire the information needed during the appeals evaluation process. The moderators from these servers typically set up separate \textit{appeals servers} that invited members who violated community policies to create a ``ticket'' or to discuss their case in a dedicated channel. 

Moderators of large servers emphasized the benefits of designing and implementing an appeals-specific server to facilitate the demanding appeals process. Appeals processes, as described in later sections, can require a complex coordination of large amounts of data and communication across various stakeholders. The appeals server consolidates all necessary information, actors, and records into one dedicated and easily accessible place. This enabled users to better understand how to appeal and allowed moderators to directly provide guidelines to the users. It also made all appeals cases transparent to other moderators. The moderation team could keep tabs and check on the ongoing cases, ensuring that any moderator or administrator could quickly jump in or intervene if the moderator who was currently handling the case was unavailable, needed assistance, or showed signs of bias. Additionally, having a separate server and using a bot to initiate the appeals process served to help protect moderators against abusive or vindictive users. When F2 had used Google Forms to receive appeals, ``\textit{spammers also used it to send [them] harassment messages through the form}.'' F1 stated that they wanted ``\textit{to ensure that moderators...feel safe when going through the ban appeal with the user without their name being shown, so they can feel comfortable in knowing that their username will not be displayed and potentially be harassed in the future if the ban appeal is rejected.}'' For these reasons, many servers created separate appeals servers as their size grew and the load on their moderation capabilities became too large. Part of this effort included bringing in moderators who had technical experience in building such systems, demonstrating social learning through knowledge transfer in a multi-level governance structure~\cite{jhaver2023multilevel}. By designing and constructing the servers to fit their needs, the community achieved much more effective and efficient appeals processes.

\subsubsection{Gathering context}
Regardless of the method used to submit an appeal, the information required to initiate the evaluation process remains consistent. When a user began the appeals process, they would be provided a form via a moderation bot or through a link. Interviewees almost all reported that the same two main questions formed the basis of their appeal forms: (1) why they were penalized, and (2) why their appeal should be accepted. Many servers also asked (3) what steps the user will take in the future to prevent the same problematic behavior from taking place (Figure ~\ref{fig:map}B, left). Most servers also provided free space to add any additional information. These questions served as starting points for understanding the situation and providing materials to help judge whether the appellant showed the right signals to be accepted back into the community. 

However, this step is just the first part of the long collection process -- after one or more moderators checked the user's message, they began gathering relevant information to build the foundation for painting the full picture of the case (Figure ~\ref{fig:map}B, right). Moderators would typically review the moderation logs from the main server to collect information such as the infraction type, the time since the infraction occurred, the problematic text or audio and its surrounding messages, the user's infraction history, and more. Automated tools were extensively used at this stage in the form of moderation bots and databases. The most frequently used bots were ModMail, Zeppelin,\footnote{https://zeppelin.gg/} and Dyno, which are general-purpose moderation bots widely used on Discord. These bots maintain logs of server messages, interactions, and moderation actions taken for the entire server. Moderators praised these bots for how they made the information-collecting process painless and described them as essential for an effective appeals process.

\begin{quote}
Zeppelin is so important for appeals in the way that it handles infractions. You're able to include screenshots and links and notes and all of that. So when you ban someone, you're able to include all of that information for whoever has to review the appeal to look at. -- F4
\end{quote}

Some servers even customized their own bots to better cater to their specific needs. P5 added and maintained a tool that allowed moderators to add comments on the infraction and the moderation decision so that the moderator who processed the appeal could easily refer back to the comment to understand the rationale behind the decision.

During this stage, the moderators emphasized that they were seeking a conversation with the user rather than a confrontation. They needed to engage with the violator to obtain relevant details and account for possible differences in perspective. If they required further information from the user or wanted to hear more from their side of the story, they would ask the user directly. Enabling two-way communication was another advantage of having an appeals server or channel instead of using a form, which makes it hard to reach out to users when questions arise. F9 stated, ``\textit{I'm generally against the form approach because you can't have a conversation. Uf you are doubtful about something that they answered in the questions, you cannot have a conversation with them to clarify.}'' On the other hand, several moderators mentioned that it also left moderators more vulnerable to harassment and badgering. Despite these dangers, moderators viewed it as crucial to evaluate the appeals within the full context of the available information, and the process of gathering this information wasn't a one-and-done step, but rather an iterative cycle that persisted throughout the entire appeals decision-making process.

\subsection{Evaluating an Appeal}
After sufficient information has been gathered, the moderators begin to assess whether the appeal should be accepted (Figure ~\ref{fig:map}C). Moderators in servers made such decisions individually or, more frequently, as a group (and in one case, as a pair). In this section, we describe the evaluation process irrespective of the number of decision-makers involved, but we discuss the collaborative aspects of the evaluation process in the next subsection.
The moderators interweaved a diverse set of context cues to reach the final decision on an appeal. This decision-making process incorporates both objective standards and subjective, experience-driven human judgment. We first describe key signals of the evaluation process that rely on definable metrics, such as time since infraction and length of the appeal. Then, we explain how moderators tacitly evaluate the behavioral cues -- such as sincerity, attitude, and likelihood of recidivism -- to make their decision. 

\subsubsection{Objective signals}

One of the most crucial yet readily apparent aspects of a successful appeal was the time gap between when the infraction occurred and when the appeal was submitted. Almost all moderators noted that the time since infraction was a defining signal as to whether the appeal would be accepted. In certain servers, moderators denied appeals that were submitted within a certain period (e.g., two weeks) after the infraction. This ``cool-down period'' effectively culled users who immediately appealed their timeouts or bans and approached the situation with frustration, anger, or indignance: P1 stated, ``\textit{People who appeal right afterward never get approved, essentially.}'' Many moderators stated that the users who appealed quickly usually did not recognize what they did wrong or seemed insincere in their apology (``\textit{If you come and immediately apologize, then it looks as if you knew what you were doing and you thought you would get away with it. But we caught you, and now you are apologizing for that. So many times it has happened that my moderators tend to just deny the appeal.}'' -- F9). However, though many servers had specific guidelines on a ``cool-down period,'' the moderators understood that this restriction should not be enforced absolutely in order to allow for the possibility of mistakes by the staff.

\begin{quote}
    It really depends on the reason they were banned, and that's why we can't really enforce a period, because users can make a mistake. But staff can also make mistakes. And we never know, sometimes a staff member may ban someone in retaliation for something, and we aren't even aware. So we can't prevent someone from requesting an appeal or asking to talk to some higher-ranked staff member. -- F8 
\end{quote}

Another important yet often clear-cut criterion was the type of harm committed. Communities drew hard lines for severe and highly harmful infractions, such as Terms of Service violations or other illegal conduct. For example, F4's community had a zero-tolerance policy for extreme offenses such as the distribution of Child Sexual Abuse Material (CSAM): ``\textit{I think it really depends on what the person did. What the infraction is, and which criteria you're willing to accept. Because if someone comes in and they're like, `I'm so sorry that I sent CSAM', I don't care that you're really sorry, you need to get banned.}'' Though most moderators held clear thresholds for universal high-level harms, such as hate speech, some standards were unique to the community norms and culture. For P8, who managed a server for Roblox gamers, the members were banned from the Discord community if they violated the Terms of Service in Roblox.  

Finally, moderators pointed out that effort was an important indicator of whether the appellant is likely to improve their future behavior. When asked how they determined the appellant's effort, the participants pointed to the length of the appeal. According to the moderators, longer appeals usually demonstrated greater remorse and indicated that the violator had considered the issue in more depth. P6 added that even if the quality of the appeal was not the best, \revision{``\textit{a longer appeal shows that maybe they have a better understanding of why they're banned and why they want to rejoin, or they put more effort into the application because they genuinely do regret what they did and want to rejoin us.}''} A few moderators noted that with the prevalence of widely available Large Language Models (LLMs), they would encounter appeals that were suspected of being written by AI. However, moderators were not significantly concerned about the impact of LLMs, as they felt that they could easily distinguish these appeals from user-written appeals: from their experience, users who submitted AI-generated appeals did not put in the time to edit the answers to fit their specific case, further demonstrating the lack of effort and making the appeal easy to reject.

\subsubsection{Subjective signals}
Ultimately, the evaluation of appeals relied on subjective moderator judgment. During the individual interviews, moderators were asked extensively about this subjective judgment process. To help with the evaluation process, communities maintained internal guidelines on how appeals should be approached to encourage a fairer evaluation process. The moderators also alluded to the concept of ``gut feeling'' in knowing which appeals should be approved. They explained that they developed this sense through experience and training, such as shadowing other moderators on their appeals cases as a junior moderator. When asked to concretize how they performed their evaluation, the moderators outlined several key criteria the appellant must meet: a clear understanding of their wrongdoing, a sincere and respectful tone of the appeal, and a commitment to change for the better. 

To help assess whether the violator grasped the harmfulness of their actions, appeals systems in all servers asked the same initial question: ``Why were you banned/penalized?'' Usually, the exact reason for the ban or other interventions had previously been provided to the user via the moderation bot when the punishment was administered, so the primary purpose of the question is not to capture the actual reason -- though a few moderators noted that banned members sometimes lost access to the message or forgot after an extended period, so it served as a reminder -- but rather as a gauge for checking whether the member recognized that their actions were wrong. However, moderators also looked for whether the appellant had gone beyond simply recognizing which rules they violated into understanding the negative impact of their behavior, stating, ``\textit{You can tell, if [the member] is simply acknowledging their mistakes or if they are actually showing remorse. You can tell just by the fact that they are apologizing in the first place and the general tone of the message. You can tell.}'' -- P4. 

Similarly, the other question asked across all appeals servers -- ``Why should we unban you/approve your appeal?'' -- aimed to test whether the member expressed sincere regret for causing harm to the other member(s) and the community. Those who refused to show such regret or apologize for their actions were rejected. P5 referred to these users as having \textit{no intention to integrate}. It should be noted that moderators often held active dialogues with the appellants to help them understand why they were penalized, explaining the rules of the community and why they must be followed. During this process, the moderators often aimed to give the users the best possible chance to reintegrate into the community. 

\begin{quote}
     I always want to give people a second chance...I'll never deny an appeal just because of what they say in the very first message. I'll ask them to take this seriously. If they don't take it seriously, then that's a huge red flag. I'll try to give them the benefit of the doubt and try to get them to be more serious. But at the end of the day, if they still continue to be [not cooperative] in their appeal ticket, then I'm gonna blacklist them. -- P5
\end{quote}

The appeals system run by P8 even included a question asking, ``Do you realize this is your last chance?'' to instill within the appellant the graveness of the situation and the importance of giving their best efforts in the appeal.

Finally, the moderators assessed whether the member's understanding and remorse would be reflected in their future behavior. Some moderators pointed out that predicting whether the user will truly behave better once their appeal is accepted is challenging, but that they develop a quite accurate ``sense'' through experience. To ensure that their decisions had been correct, some moderators kept a close eye on the appellants for a while after they were ``released'' from their term to make sure that they did not recidivate.

\subsection{Collaborating on Appeals Decision-Making}
Though evaluating the appeal is already a time-consuming task, the majority of the servers chose to handle all appeals as a group throughout the whole process rather than leaving the decision to an individual moderator. We discuss in detail how community moderators coordinate collective decision-making and how leveraging group-based decision-making upholds fairness, consistency, and due process. 

\subsubsection{Collaboration process}
In many servers, the collaboration between moderators begins as soon as an appeal is submitted. Even in servers where most appeals cases were handled primarily by an individual moderator, a group discussion could be triggered by a moderator looking for a second opinion or facing difficulties with evaluating a special case that had no clear precedent. Often, other moderators who raised an issue with how the moderation was being handled~---~such as a moderator giving out what was perceived as a biased decision~---~or who hoped to assist another moderator, could also launch a discussion. The moderator group dynamics for appeals were often hierarchical. For example, in P4's servers, senior moderators and administrators would be in charge of handling more serious cases. In nearly all servers, the server administrators held the final say in how an appeal should be handled in case of disagreements or special circumstances. 

During the discussion, all involved moderators have access to the relevant information gathered during the initiation process (Figure ~\ref{fig:map}D, right). This information might sometimes be shared in a moderator group channel via links or screenshots by the individual moderator in charge of collecting all evidence. If need be, the group would ask the moderator who gave out the original penalty (if given by a human rather than a bot) or the moderator who has been handling the case to share more information and impressions. The groups would then discuss the case after assessing all the information given. Discussion points would include aspects such as appeal quality (the overall quality and thoughtfulness of a user's appeal), proportionality (whether the scope and duration of the ban are appropriate given the user's actions and appeal efforts), the possible consequences of approving or denying an appeal, precedent cases, and any possible bias involved in the decision making. The participants mentioned different types of biases that could be discussed in the group. The bias could be (1) case-specific -- viewing certain cases as inherently more or less severe based on personal beliefs or experiences, (2) appellant-specific -- holding views towards certain users based on their interactions or reputation, or (3) policy-specific, in which a moderator may be perceived to be using the case as a way to express a difference of opinion about certain community policies.

\subsubsection{Handling disagreements}
Through this process, the moderator teams sometimes encountered disagreements (Figure ~\ref{fig:map}D, middle). When asked how they work through differences in opinions, most participants stated that it was rare to have disagreements as moderators within a given community typically shared similar values and attuned their judgements over time. However, though rare, it was also possible that certain cases took a long time to reach a final decision or that tensions escalated between moderators into arguments. Some servers avoided lengthy stalemates and conflicts by having all moderators vote for each appeal decision.

While sometimes these disagreements could evolve into unpleasant conflicts, they also became opportunities for moderators to broaden their perspectives. F2 recounted that disagreements could occur when the appeal was poor in quality, but could be counterbalanced by the low level of harm and lack of previous history. This would reveal which factors moderators within the server prioritized and helped realign the internal standards. P3 in particular emphasized their penchant for healthy disagreements: 

\begin{quote}
There is a lot of disagreement. I try to encourage disagreements. Disagreements help have a healthy conversation, where we can look at multiple viewpoints. I don’t disagree for the sake of disagreeing, of course, and if no one brings it up, then we don’t talk about the viewpoint. But then there might be more insight from someone else, or you might change their mind by talking about these different perspectives.~--~P3
\end{quote}

In cases in which no consensus could be reached, almost all servers gave the final decision power to the server administrator (Figure ~\ref{fig:map}D, left). Several servers described a structure in which the administrator would not be actively involved in the day-to-day appeals decisions but could be relied on to be the voice of authority in difficult cases.

\subsubsection{Achieving Fairness}
According to the interviewees, the core tenet of the appeals process was fairness, and this was often achieved by involving multiple moderators. \revision{Moderators emphasized that community appeals systems uphold fairness through their procedural design.  Participants' concepts of fairness within appeals systems included consistency of actions taken between cases, removal of personal bias in decision-making processes, and accountability for each decision. These principles were achieved by using a consistent, structured process, including defining formal appeals guidelines for moderators to follow.} F3 emphasized, ``\textit{I think the most important thing was to establish a structured process to ensure a standardized process. It's very important to us that banned users are treated fairly.}'' Moderators put great importance on building multiple, layered mechanisms to create an impartial system for all users. Moderators adhered to structured and standardized processes for handling appeals, often documented through formal guidelines. Some servers even excluded moderators who had given the original penalty from participating in an appeals process (``\textit{A lot of times, moderators don't like going back on their decisions}'' -- P3) or excluded moderators who had personal relationships with the appellant in order to remove personal preferences and prejudices from the review process. P7's moderator team also added an extra layer of user protection by allowing users to report staff who conducted themselves inappropriately, including making biased decisions during appeals.

Transparency among the community moderators was crucial to ensuring fairness. The appeals server gave access to every appeal case to all moderators at any point in time. Certain servers forbade users from contacting individual moderators privately, for both the moderator's safety and to make all interactions public to the staff. While users did not have a direct window into the deliberations over their appeal, the moderators often provided reasons behind their decisions to the appellants, especially if it was a rejection. This included referring the user to existing community rules and how they had violated those rules. By keeping each other accountable and all records visible, the moderators strived for consistency and impartiality.

\subsection{Values of Appeals Systems}
As shown by Atreja et al. and reiterated by many of our participants, designing, building, and executing an appeals system demands much from moderators who are already overworked~\cite{atreja2024appealmod}. Yet, many communities decide to invest precious time and resources into building out sophisticated appeals systems because they view appeals systems to be valuable for their communities. We describe how enabling appeals encourages both individual and community growth. 

\subsubsection{Encouraging personal growth}
The very first answer that almost all moderators stated when asked about the value of having an appeals system was that it provided a second chance to the members. One participant mentioned that for some people, especially young members, the community is a huge part of their social life. Several moderators empathized with the members and aimed to work \textit{with} the member in their best interest during the appeals process, rather than as a strict judge or enforcer of rules.

The appeals system also left the door open to those who returned to the community improved and ready to be better. Several moderators recounted cases in which a banned member would appeal to be let into the community after a long time had passed. P4 stated, ``\textit{We've had cases where a person who was banned years ago came back, saying, `I was young and I've changed as a person since then.' In those cases, we want to have appeals open for them.}'' These members, who had grown since their infraction, were also thought to be deserving of another chance. 

Most notably, the appeals system spurred members to reconsider their actions and reform their behavior at a fundamental level. In many cases, it was evident that some had simply never learned how to properly conduct themselves, and appeals systems served as an educational tool for becoming a better community member. The benefits of educating users are also propagated to the community. F5 told the story of a member who had been banned due to unruly behavior but went through a successful appeals process. This, in turn, allowed them to grow and motivated the member to actively participate in the community, eventually joining the staff and adding their perspective to the process: ``\textit{I'm so glad they're on the team because there's so much more value that they bring just by being able to empathize on that level.}''  

However, many moderators also expressed that certain people simply could not be expected to do better, no matter how many chances or explanations were given. Moderators learned through experience how to distinguish these individuals, and some viewed time wasted on such cases as one of the biggest hardships in the appeals process.

\subsubsection{Strengthening communities}

Finally, \revision{speaking from their personal experiences interacting with users during and after appeals processes,} \revision{moderators felt that} appeals systems also added to community integrity and safety. Even if the appeals system did not benefit every individual within the community, \revision{moderators felt that} having one helped foster a sense of belonging \revision{in the} community \revision{for those who went through the process, and also a belief that the community practiced justice in moderation decisions}. F5 stated that especially in smaller communities, nurturing afforded through the appeals systems ``\textit{really [fosters] a sense of belonging. If you're able to show that bit of investment, people do feel valued and they do understand that they can be heard, they can change, and they have a lot more choice as to what they may originally think.}''

Appeals systems also offered insights into the potential values of those who appealed, especially those who displayed a lot of effort in their appeal and indicated their passion and dedication to the community. These individuals could be valuable assets to the community's long-term growth and culture. P4 pointed out how servers were extrinsically motivated to retain high membership and activity. Appeals systems increase the chance of retaining high engagement and participation. At the same time, moderators warned that one had to be careful in balancing this external need as lenient standards quickly unraveled the credibility and fairness of the appeals system. Additionally, appeals systems helped filter out users who may pose a continued risk to the community, as it also gave clear signs to which members were simply not redeemable.

%% file: sections/05_discussion.tex
\section{Discussion}

By mapping the community appeals process in Discord, we find that the process demonstrates how communities successfully incorporate communal needs and values in their socio-technical designs for complex, human-driven moderation processes. The moderators adaptively built appeals systems with \revision{supportive technical tools} as their communities grow. However, even as the communities became larger, moderators maintained procedural fairness in evaluating the appeal through transparent and collaborative evaluation. Through this design, appeals systems educate and reform members to be more responsible participants in the community.

\revision{In this section}, we discuss how the example of Discord community appeals systems show paths to integrating community values into systematic moderation processes. We first consider appeals as a feasible approach to rehabilitative practices in community moderation. Then, we frame how moderators can be supported in community-driven moderation processes through adopting tools and education. \revision{We also point to how moderators use appeals systems to establish legitimacy through their pursuit of fairness.} Finally, we highlight the importance of local processes in supporting platforms and what lessons platforms can learn from appeals within the multi-level governance framework.

\subsection{Adopting Rehabilitative Moderation Practices}

A growing body of work in content moderation design is applying \revision{rehabilitative~\cite{cullen2013reaffirming, seigert2007rehabilitation, kou2024peer} and restorative principles to online moderation practices~\cite{petersonsalahuddin2024repairing, xiao2023restorative, hasinoff2022subsidiarity, kou2021punishment, schoenebeck2021justice, doan2025apolobot}}. Unlike punitive justice systems, which focus on content-based penalties, restorative justice prioritizes individuals. According to prior work, one major limitation in adopting restorative justice into existing online communities is the amount of resources and effort required, which often adds considerable strain to the community and its volunteer moderators acting as mediators~\cite{xiao2023restorative}.

Appeals systems serve as an intermediate yet realizable step toward restorative principles as a rehabilitative justice-based approach to moderation. A major strength of the appeals processes we describe in this paper lies in their relative expedience compared to restorative justice practices that require the involvement of multiple stakeholders. \revision{Xiao et al., who examined the application of restorative justice principles for cases of interpersonal harm in online gaming communities, encourage communities and platforms to ``design restorative justice mechanisms based on their resources, culture,
and socio-technical affordances''~\cite{xiao2023restorative}. Discord community appeals systems embody this precise idea, echoing their suggestions of involving a few select stakeholders or practicing moderation with restorative justice values.} The reported success of \revision{Discord community} appeals systems shows that the road to achieving restorative justice does not have to be all-or-nothing; communities can still benefit from partial adoption of restorative justice principles or from interventions that draw inspiration and motivation from restorative justice but do not yet fully implement restorative processes. From the implementations of the Discord servers' appeals system, we derive socio-technical design considerations, such as engaging in timely yet constructive conversations and ensuring safety from secondary harm. These considerations can support incorporating other aspects of restorative justice, like initiating dialogue between the victim and the perpetrator or inviting the community into the governance process~\cite{mcccold2000holistic, zehr2015little}.

Additionally, appeals systems can be viewed as an important tool for encouraging \revision{users' trust in community governance. Moderators believed that constructing a legitimate and just governance system is important to maintaining the community's trust.} \revision{We further discuss how moderators establish the legitimacy of community-led appeals systems in Section ~\ref{legitimacy}.} This trust is conducive to building a healthy community, as it encourages users to engage constructively with the platform's rules and values. \revision{Appeals processes align with values of dialogue, education, and rehabilitation, which has shown to be particularly effective in maintaining this trust~\cite{schoenebeck2021justice, suzor2019transparency}}, as they demonstrate a commitment to fair treatment and the well-being of users. Through this lens, appeals systems \revision{can} contribute to both procedural fairness and the long-term sustainability of online communities by promoting restorative principles and supporting user reintegration. \revision{However, we also recognize that most users of a community do not partake in appeals processes and thus may remain unaware of their mechanisms and impact. Our understanding of the effect of appeals systems will benefit from further exploration how community appeals systems shape Discord users' trust in governance.}

\subsection{\revision{Supporting} Effective Community Governance through Moderator Education and Tool Building}

Though community appeals systems promise significant social value, building and maintaining appeals processes in communities can increase the workload for already-overburdened moderators~\cite{seering2019moderator}. Community moderation is often human-driven; appeals, especially so. Appeals systems and other moderation tasks are helmed by human-intensive processes to uphold due process and community interest. While previous works have highlighted that human-driven, context-aware processes may be slow and inconsistent~\cite{jiang2023tradeoff}, our results show moderators \revision{attempt} to address these challenges by 1) maintaining consistency through clear definitions, training and educating moderators in how to manage fair appeals processes and 2) expediting context sharing by adopting customized support tools.

We observe that moderators take measures to ensure fair and consistent decision-making. To maintain procedural consistency, many communities adhere to documented guidelines and share best practices. Even when performing subjective evaluation, community moderators use certain objective signals (e.g., time since infraction) to explain and motivate their decisions. This echoes findings from Wohn et al. on how moderators on Twitch profile users based on extensive evidence collection and analysis of recurring behavioral patterns~\cite{wohn2021profiling}. It should be noted that hands-on experience is still imperative in learning to manage appeals processes, as with many other tasks in moderation; moderators grew their ``gut feeling'' by recognizing emergent patterns from their experience. Furthermore, collaborative and hierarchical decision-making structures --- in which more senior and experienced moderators take charge of difficult cases --- teach moderators to be accountable and transparent. These checks and balances show that, though appeals systems rely on human judgment, \revision{community moderators strive to} uphold consistency and procedural fairness in the judicial process through education and collaboration. 

In addition to judicial measures, community moderators use technical means to support context-sharing. As demonstrated in many servers, bots can expedite the arduous task of information gathering and sharing. This is especially important when the information may be distributed across multiple channels and even outside of the community. Also, tools can provide more opportunities for constructive and meaningful conversation with the users by enabling moderators and users to discuss resolutions and community guidelines more directly within the shared context. Similar to the findings by Atreja et al.~\cite{atreja2024appealmod}, we also saw that moderators felt safer with bots serving as a buffer from abusive users. By focusing on the use of tools to enhance interpersonal interactions rather than replacing them, communities can foster stronger relationships between the governing body and its members.

Such processes are difficult to achieve without providing community moderators with the tools and knowledge. Thus, we urge platforms to take on a more active role in facilitating these systems. Discord, for instance, offered Moderator Academy Exams, which were designed to teach and test moderators on recommended practices~\cite{discord2023academy}. This example shows how platforms can provide more structured tools and resources, leading to a more sustainable and cooperative governance model~\cite{seering2022pride}.

\subsection{\revision{Establishing Legitimacy in Community Governance}} \label{legitimacy}
\revision{Moderators performing community appeals work demonstrated firm belief in the procedural fairness of the process, emphasizing consistency across users, elimination of bias, and creation and enforcement of guidelines. We find that these core concepts of fairness in appeals systems as described by moderators harken to how \textit{legitimacy} can be established in a community-based governance structure. Zelditch describes legitimacy as ``a process that brings the unaccepted into accord with accepted norms, values, beliefs, practices, and procedures''~\cite{zelditch2001processes}. Legitimacy impacts perceptions of justice, as the lack of legitimacy in the system leads it to be perceived as ``less fair'' for those disadvantaged by the system~\cite{melamed2012effects}. Thus, for governing bodies, such as community moderators determining the outcome of an appeal, it is crucial that its decision-making process is perceived to be legitimate.} 

\revision{Our results show that community moderators of Discord appeals systems attempt to strengthen the perceived legitimacy of community governance. Just like in modern civic systems, which is described to be ``a more or less polycentric system in which a variety of actors are engaged''~\cite{denters2010sage}, online governance involves a complex network of various stakeholders. Haesevoets et al. applied a three-component model of legitimacy (\textit{input}, or inclusivity and responsiveness, \textit{throughput}, or transparency and integrity of procedural implementation, and \textit{output}, or governance performance)~\cite{bekkers2016legitimacy} to understand how perceived legitimacy can be multifaceted across different stakeholders and decision-making modes~\cite{haesevoets2024towards}. Similarly, we position our findings of how moderators of Discord community appeals systems describe the community appeals process within this framework --- moderators often described that their appeals systems aimed to be inclusive and accessible (input), transparent, impartial, and accountable (throughput), and efficient and effective (output). As Bekkers and Edwards point out, the decision-making procedure, not its content, forms the basis for legitimacy~\cite{bekkers2016legitimacy}; community moderators work to implement and maintain this procedure.}

\revision{We note that this work contributes to exploring how legitimacy is shaped in community governance, while also acknowledging its limitations and calling for further probe into the perceived legitimacy and fairness by other stakeholders. Though previous work compared users' perceived legitimacy of content moderation processes by contractors, algorithms, expert panels, and impartial digital juries~\cite{pan2022comparing}, there has been a relative lack of research on how community moderator-driven governance is perceived to be legitimate. Unlike external governing bodies, community moderators are deeply embedded within the very structure they regulate, which presents unique challenges and opportunities for legitimacy. On the other hand, our work provides a moderator-centric perspective on governance, limiting our understanding of how these systems may impact users. Future research can explore user perspectives on the legitimacy of community-driven moderation systems and how their participation can enhance it. Additionally, platforms may consider mechanisms similar to those used by communities to strengthen their legitimacy, such as enhanced transparency and participatory governance. Prior work suggests that platforms often fall short in upholding legitimacy, as they sometimes fail to adhere to consistent rule-of-law principles~\cite{suzor2018digital}. Examining legitimacy across different levels of governance can help define more effective governance models.}

\subsection{Designing Multi-level Governance through Community-driven Processes}

Most of the appeals systems we have discussed operate at the local server level and the question remains, what are the lessons platforms can learn from this process? Currently, safety in major platforms is frequently evaluated in terms of how quickly and consistently harmful content is removed rather than by how well platforms facilitate positive growth among users and communities. Thus, the metrics that platforms prioritize seem to discourage adopting human-driven, context-aware moderation practices, as they contradict a platform's vested interests -- namely, the intersection of efficiency and compliance with relevant regulations. However, we see that the implementation of a user-centered, community-driven appeals system can lead to increased engagement and commitment from users. From a different perspective, a user who has lost trust in the platform is a user lost forever. Keller emphasizes that by chasing immediate and easily quantifiable measures, such as time-based metrics that promote quantity over quality, negative consequences ripple to all levels of governance~\cite{keller2024compliant}. Though better integrating users and communities into a multi-level governance architecture for platform governance may seem like a radical goal, it may actually be better aligned with platforms' short and long-term interests than current more centralized approaches. 

\revision{By definition, polycentricity requires the presence of multiple independent decision-making centers~\cite{ostrom2017polycentric}. Previous work on polycentricity in traditional physical forms of governance defines polycentricity into Type I, in which authority is dispersed among a limited number of non-overlapping centers and levels, and Type II, in which governance is more fluid, overlapping, and changes on-demand~\cite{bache2016multi}. Discord consists of various communities (servers) operating with a significant degree of independence and flexible, adaptive governance, with governance powers overlapping in many ways with platform governance processes. Each server can have its own rules, norms, and culture while sharing resources and membership across different servers, making it resemble a Type II model where the hubs are interconnected without one central authority dictating all the interactions. They particularly embody contextualization and value-driven operationalization even among community-centered social media platforms such as Reddit, thanks to the robust set of social and technical affordances available in defining the interactional ``space'', where users collectively shape and enforce norms and practices~\cite{smit2024digital}. The structured Discord community appeals system represents an example of how distributed governance processes can lead to more adaptive and responsive forms of community regulation. Smitt et al. state that strong senses of ownership, belonging, and shared responsibility emerge from the``space''~\cite{smit2024digital}. This is reflected in moderators' perspectives that managing a rehabilitation-focused appeals system fostered a stronger sense of belonging for users who successfully completed the process.}

\revision{We also consider how Discord appeals systems are positioned within hierarchical and networked models of multi-level governance. Though community moderators have autonomy and authority for community-specific moderation decisions, the Discord platform holds authority in enacting platform-wide moderation policies and handling serious cases. Moderators respect the authority of platforms, such as abiding by the Terms of Service in their appeals moderation decisions. Even within the server exists a similar hierarchical structure in the form of an admin, who would make a final decision in contentious cases. We also observed the networked sharing of information and resources in establishing appeals systems, such as inviting personnel who have experience in setting up appeals systems in other communities. We emphasize that one type of multi-level governance model is not inherently more or less effective than others. For example, while WhatsApp groups retain autonomy through end-to-end encryption limiting access from platform operators~\cite{jhaver2023multilevel}, Discord delegates autonomy to middle levels of governance by providing custom channel configurations, assignable roles for different levels of access, and the option to integrate third-party bots. However, we hold that there is value in polycentric governance for online platforms. Hoogeh and Marks found that ``a common element across [polycentric governance literature] is that the dispersion of governance across multiple jurisdictions is both more efficient than and normatively superior to central state monopoly''~\cite{hooghe2010types}}.

Facilitating the adoption of community-level appeals systems can thus be seen as a practical step in the development of context-sensitive yet scalable governance. By applying rehabilitative justice principles, \revision{Discord moderators attempt to foster healthier communities.} This transition is not without challenges, particularly in terms of standardizing policies and allocating resources. Yet, we observe through Discord community appeals processes that it is possible to achieve rehabilitative principles at scale. Platforms can enable the development of these systems at a community level as part of a broader multi-level approach to governance.

However, we note that expanding this process at the platform level does not mean replacing existing local systems. Community-level governance is inherently -- and perhaps ideally -- not infinitely scalable. The tension of scalability and context-sensitive and individualized moderation, referred to as ``subsidiarity'' by Hasinoff and Schneider~\cite{hasinoff2022subsidiarity}, emphasizes that local social units in larger systems should reserve meaningful autonomy. In this vein, the platform-level systems should operate alongside community-level governance through active support of local processes, the exchange of ideas and affordances, and the sharing of jurisdiction and autonomy, rather than attempting to supplant community efforts.

%% file: sections/06_conclusion.tex
\section{Conclusion}
Online governance has long been dominated by a view that platform-driven and community-driven governances are inherently divided in their interests and practice. However, this study highlights through a qualitative analysis of community-level appeals systems in Discord that community-led \revision{moderation} can achieve scalable and consistent moderation that also incorporates community-driven designs. By focusing on the human-driven appeals processes within Discord communities, we observed how moderators use technical resources and collaboration structures to create fair and efficient appeals processes. This research contributes to the understanding of multi-level governance by illustrating how community-led systems fill important gaps, especially in promoting restorative justice and reinforcing community values. Ultimately, our findings underscore the importance of integrating flexible, user-driven governance into larger platform moderation frameworks to ensure a more equitable and effective approach to addressing harmful behaviors online.

%% file: sections/07_appendix.tex
\section{Appendix}
\label{A}
We used semi-structured approaches for both focus group and individual interviews. Researchers adhered to core questions and themes but adapted the flow as needed to probe emerging topics. The preliminary questions are listed below.

\subsection{Interview Questions for Focus Group Interviews}

\begin{enumerate}
\item What types of appeals have you encountered? 
\item What are the main reasons people typically appeal bans?
\item What criteria do you use when evaluating appeals? 
\item When are appeals handled individually versus in discussions with other moderators?
\item What server characteristics affect the appeal system used?
\item What tools, bots, or other resources do you find most effective for managing appeals?
\item What are the key lessons you’ve learned over time in handling appeals?
\item What motivates your community to have an appeals system?
\end{enumerate}

\subsection{Interview Questions for Individual Interviews}

\subsubsection{Introduction and Ice Breaking}
\begin{enumerate}
  \item Could you briefly describe your experience as a moderator? 
  \item What type of servers do you moderate? 
\end{enumerate}

\subsubsection{Experiences in Handling Appeals}
\begin{enumerate}
    \item What types of appeals have you encountered?
    \item What is your specific role within the appeals system?
    \item What tools do you use to manage appeals and why did you choose them?
    \item What are the biggest challenges you face in handling appeals?
\end{enumerate}

\subsubsection{Judging Appeals}
\begin{enumerate}
    \item Please walk us through the process for evaluating an appeal.
    \item What qualities do you look for in a successful appeal? What are the green or red flags?
    \item What are some cases that are especially difficult to evaluate?
    \item How have you developed your evaluation skills?
    \item How do you assess if the user is likely to improve in the future?
    \item Have you ever followed up on these cases? What have you seen?
\end{enumerate}

\subsubsection{Collaboration Process}
\begin{enumerate}
    \item When are appeals handled by an individual versus by a group?
    \item How do the group discussions proceed?
    \item How do you handle disagreements between moderators?
    \item How do you know that the moderators are consistent and fair in their rulings?
\end{enumerate}

\subsubsection{Value and Future of Appeals Process}
\begin{enumerate}
    \item What motivates your community to have an appeals system?
    \item What tools do you think might be helpful in handling the appeals process?
\end{enumerate}